\documentclass{aa}
\usepackage{amsmath}
\usepackage{psfig}
\usepackage{astrobib}
\usepackage{journals}

\begin{document}

\title{The Nature of the 10 Kilosecond X-ray Flare in Sgr A*}

\author{Sera Markoff\thanks{Humboldt research fellow}
\and Heino Falcke\and Feng Yuan \and Peter L. Biermann}
\institute{Max-Planck-Institut f\"ur Radioastronomie, Auf dem H\"ugel
69, D-53121 Bonn,
Germany} 

\titlerunning{X-ray Flare in Sgr A*}
\authorrunning{Markoff et al.}
\offprints{smarkoff@mpifr-bonn.mpg.de}

\date{A\&A Letters, in press}

\abstract{The X-ray mission {\em Chandra} has observed a dramatic
X-ray flare -- a brightening by a factor of 50 for only three hours --
from Sgr A*, the Galactic Center supermassive black hole.  Sgr A* has
never shown variability of this amplitude in the radio and we
therefore argue that a jump of this order in the accretion rate does
not seem the likely cause.  Based on our model for jet-dominated
emission in the quiescent state of Sgr A*, we suggest that the flare
is a consequence of extra electron heating near the black hole. This
can either lead to direct heating of thermal electrons to $T_{\rm
e}\sim6\cdot10^{11}$ K and significantly increased synchrotron-self
Compton emission, or result from non-thermal particle acceleration
with increased synchrotron radiation and electron Lorentz factors up
to $\gamma_{\rm e}\ga10^{5}$. While the former scenario is currently
favored by the data, simultaneous VLBI, submm, mid-infrared and X-ray
observations should ultimately be able to distinguish between the two
cases.   }
\maketitle

\section{Introduction}

Sgr A*, the compact radio core at the center of our Galaxy
\cite{ReidReadheadVermeulen1999,BackerSramek1999a}, has been
perplexing modelers since its discovery \cite{BalickBrown1974}.  In
contrast to nearby LLAGN \cite{Ho1999}, Sgr A* was until recently only
positively detected as a radio source.  Its mass is determined at
$2.6\cdot10^6 M_{\sun}$ within $\sim 0.01$ pc
\cite{HallerRiekeRieke1996,EckartGenzel1996,GhezKleinMorris1998} and
its integrated radio luminosity has remained steady within a factor of
two \cite{ZhaoBowerGoss2001}, at $\sim 10^{-9}$ orders of magnitude
less than its corresponding Eddington luminosity.  All models to
explain the radio emission so far have focused on radiative
inefficiency as the primary explanation for this dimness, and are
comprised mainly of accretion/inflow solutions
\cite{MeliaLiuCoker2001,NarayanMahadevanGrindlay1998}
outflow solutions (\citeNP{FalckeMannheimBiermann1993};
\citeNP{FalckeMarkoff2000}, hereafter FM00) and combinations thereof
\cite{YuanMarkoffFalcke2001}.  A recent review of Sgr A* can be found
in \citeN{MeliaFalcke2001}.

Recently, Sgr A* was finally detected in the X-rays by {\em Chandra}
\cite{Baganoffetal2001} with a rather soft spectrum.  
During the second observational cycle, \citeN{Baganoffetal2001b}
detected an X-ray flare lasting about 10 ks and with a peak
luminosity $\sim 50$ times higher than the quiescent state
\cite{Baganoffetal2001}.  The averaged flare spectrum after taking
into account dust scattering is best fit with a power-law (spectral
index $\alpha\sim0.3$), which is significantly harder than that of the
quiescent state ($\alpha\sim1.2$).  The longest time scale (10 ks)
corresponds to $\sim 390 r_{\rm s}$ where $r_{\rm
s}=2GM_\bullet/c^2$ is the Schwarzschild radius, which argues against thermal
bremsstrahlung from the outer radii, e.g. from a standard Advection
Dominated Accretion Flow (ADAF;
\citeNP{NarayanMahadevanGrindlay1998}).  The smallest time scale in the
flare is roughly $600$ s, suggesting activity at scales of $\sim 20
r_{\rm s}$, which means the flare originated close to the central
engine.

The variability and the spectral index of Sgr A* in the X-rays are
consistent with synchrotron self-Compton (SSC) from the innermost
regions near the black hole, e.g., the nozzle of a jet (FM00;
\citeNP{YuanMarkoffFalcke2001}) or a magnetic dynamo within the
circularized accreting plasma \cite{MeliaLiuCoker2001}.  In this
picture, the X-rays are inverse Compton up-scattered synchrotron
photons from the so-called submm-bump
\cite{SerabynCarlstromLay1997,FalckeGossMatsuo1998}. Since the
submm-bump is thought to be produced close to the black hole, very
short time scale variability (several hundred seconds) was already
predicted (FM00).  In the following we would like to explore the
various scenarios which could lead to a dramatic X-ray flare within
the jet model.

\section{Models}
We start with our basic jet emission model
(\citeNP{FalckeBiermann1999}; FM00), consisting of a conical jet with
pressure gradient and nozzle. The parameters in the nozzle for the
quiescent state are determined from the underlying accretion disk as
described in \citeN{YuanMarkoffFalcke2001}, and as summarized below.  All
quantities further out in the jet are solved for using conservation of
mass and energy, and the Euler equation for the accelerating velocity
field. We take the distance to the Galactic center as $d_{gc}=8.0$
kpc.

Clearly, in order to produce an X-ray flare, one or several parameters
had to have suddenly changed in Sgr A*. In Figs. 2 and 3 in FM00, we
showed how the radio and X-ray spectra in the jet model change if one
changes the magnetic field -- a similar result would be expected for
an increase in particle density -- or the electron temperature by a
small amount. The former would be expected for an increased jet power
or accretion rate, which would result in simultaneous flaring at all
frequencies with little change in spectral index.  In the latter
scenario, however, the X-rays flare much stronger with a hardening of
the spectrum, because SSC is very sensitive to changes in electron
energies.  This type of fast heating could in principle occur via
instantaneous transfer of energy from the magnetized plasma in the
accretion flow to the radiating particles, e.g. as would be expected
from the sudden discharge of energy in magnetic flares through
reconnection (e.g., \citeNP{Biskamp1997}).

On the other hand, we know that non-thermal particle distributions are
quite common in jets in AGN and X-ray Binaries (XRBs), leading to the
appearance of optically thin power laws in the spectra.  Observations
of jets in XRBs (e.g., \citeNP{Fender2001}) and some AGN (e.g.,
\citeNP{MeisenheimerYatesRoeser1997}) seem to hint at a common type of
power law with typical spectral index of $\alpha\sim0.6-0.8$.  While
the exact mechanism is not yet firmly established, and reconnection
may also contribute, first order diffusive shock acceleration leads
more naturally to an electron distribution with the index $p\sim2-2.6$
depending on the shock compression ratio ($\frac{dN}{dE}\propto
E^{-p}$, see e.g., \citeNP{JonesEllison1991}).  Such accelerated
particles would result in a significant increase of optically thin
synchrotron emission, with spectral slope $\alpha=(p-1)/2$.

In the following we therefore explore three scenarios for the origin
of the X-ray flare: increased jet power or accretion rate, increased
heating of relativistic particles, or sudden shock acceleration
of the particles.  We will refer to these three models as the
$\dot{M}$-flare, the $T_{\rm e}$-flare and the shock-flare,
respectively.

\section{Results}

Since no simultaneous radio or mid-infrared (MIR)observations are available
we include in our figures an ``upper radio envelope'', showing the
highest flux ever detected at each radio frequency in long-term
monitoring of Sgr A* with the VLA \cite{ZhaoBowerGoss2001}.  While it
is possible that this type of X-ray flare is so rare that it was never
before captured by radio observations, it seems statistically unlikely
given the huge radio database compared to only two cycles of Chandra
observations. This argument does not hold for the poorly sampled data
at other wavelengths and we only consider single-epoch measurements
which most likely only reflect the quiescent Sgr A* spectrum.

\begin{figure*}[t]
\centerline{\hbox{\psfig{figure=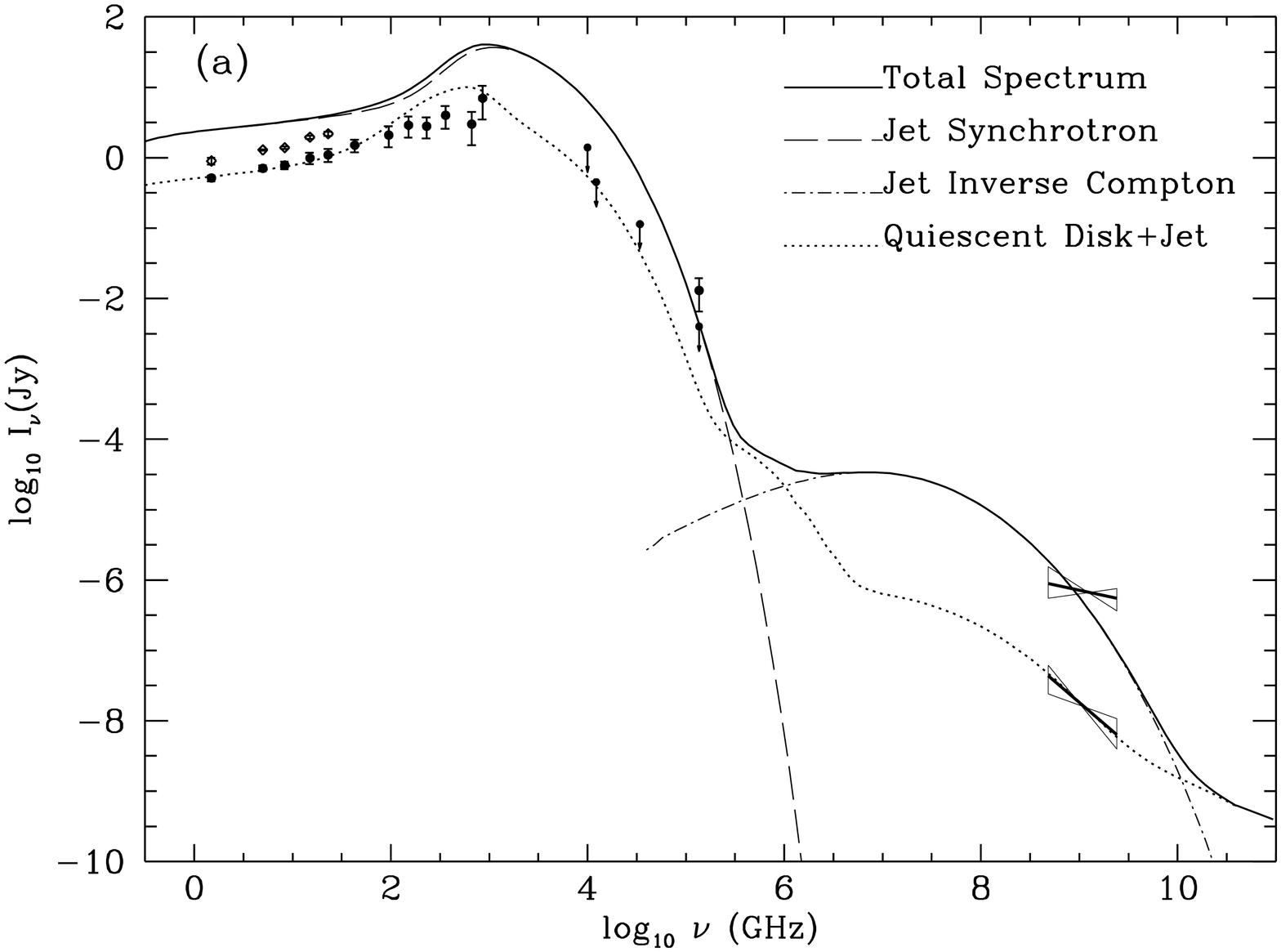,width=.49\textwidth,angle=-90}\hfill\psfig{figure=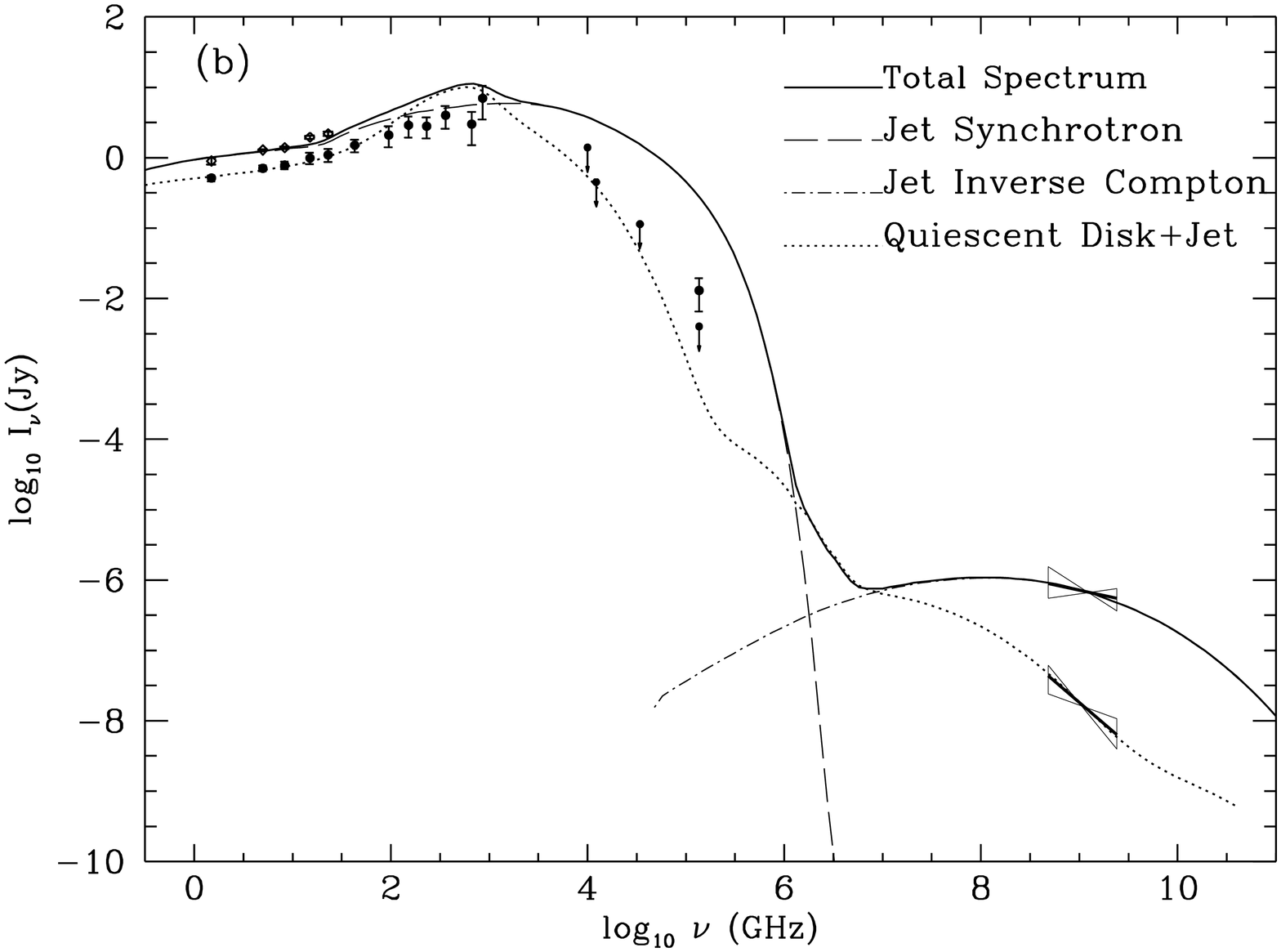,width=.49\textwidth,angle=-90}}}
\caption[]{\label{sgra_ssc} Fit of the jet model {\bf (solid line)} to the flare data of
\citeN{Baganoffetal2001b} (a) with the $\dot{M}$-flare, raising the
jet power by a factor of $\sim 3$ , and (b) with the $T_{\rm e}$-flare
by raising the temperature of the electrons by a factor of 3, compared
to the quiescent jet and disk model {\bf (dashed line)}.  The radio
data and IR upper limits are from the data set compiled and presented
in Melia \& Falcke 2001, where IR data are from single-epoch
observations only.  The upper radio points show the highest flux
detected at that particular frequency, as compiled by
\citeN{ZhaoBowerGoss2001}.  The lower X-ray data show the quiescent
state spectrum of
\citeN{Baganoffetal2001}.}\end{figure*}

The effects of the $\dot{M}$-flare and the $T_{\rm e}$-flare can be
modeled simply by changing the jet power and temperature,
respectively, in our published models (FM00,
\citeNP{YuanMarkoffFalcke2001}).  We assume that the jet carries away
a fixed fraction of the accretion energy $\dot{M} c^2$, and that this
energy is divided evenly between the kinetic energy carried by the
cold plasma, and the internal energy carried by the magnetic field and
hot electrons.  Once the electron temperature $T_{\rm e}$ in the
nozzle is fixed, assuming a Maxwellian distribution, the jet nozzle
density $n_0$ is determined via approximate equipartition from the
magnetic field $B_0$.  In the quiescent state, the relevant parameters
for our most recent fit are $T_{\rm e}=\sim2\cdot10^{11}$ K,
$n_0\sim9\cdot10^5$ cm$^{-3}$ and $B_0\sim20$ G.  Fig.~\ref{sgra_ssc}
shows the prediction for a) the $\dot{M}$-flare, with jet power
($\propto B^2$) raised by $\sim 3$ via increasing the jet nozzle
magnetic field $B_0$ to $\sim35$ G while holding $T_{\rm e}$ fixed
(which in turn increases $n_{\rm 0}$ by $\sim 3$ to $\sim 3 \cdot
10^6$ cm$^{-3}$) and b) the $T_{\rm e}$-flare for $T_{\rm e}$ raised
by a factor of $\sim 3$ to $T_{\rm e}\sim6\cdot10^{11}$ K, while
holding $n_0$ and $B_0$ fixed. The parameters were chosen to match the
amplitude of the X-ray flare data, shown with its error box as well.
For comparison we show in the figure also the quiescent jet+disk
spectrum.

As expected, the $\dot{M}$-flare strongly over-predicts the radio flux
by a large factor.  In fact, such a huge flare in the radio has never
been reported and in addition, the spectral index is far too steep.
The $T_{\rm e}$-flare fares much better: the predicted radio flux is
close to already observed radio flare maxima and the X-ray spectrum
becomes very hard during the flare.  The model also predicts
significant brightening in the MIR range during the
X-ray flare event, due to the shift of the submm-bump to higher
frequencies, which should exceed currently available non-simultaneous
MIR/NIR limits. In contrast to the radio, the MIR regime has not been
sampled well enough to decide whether such flares exist. However,
\citeN{GenzelEckart1999} and \citeN{SerabynCarlstromLay1997} report
observations where Sgr A* could have been detected during a brief
period with unusually high flux densities at 350 $\mu$m and 2.2
$\mu$m. Clearly, this needs to be confirmed and reassessed in light of
the new X-ray observations.

The shock-flare scenario requires more discussion, as it involves the
effects of diffusive shock acceleration in the jet.  This has been
done already by \citeN{MarkoffFalckeFender2001}, where the scaled
version of the jet model previously used to explain Sgr A*
(\citeNP{YuanMarkoffFalcke2001}; FM00) has successfully been applied
to X-ray binaries in the low/hard state by including shock
acceleration. Because the low/hard state is characterized by a very
faint, possibly ADAF-like accretion disk as in Sgr A*, the ambient
photon field is not strong enough to result in significant inverse
Compton (IC) cooling, allowing shock accelerated electrons to achieve
rather high energies. 

Following \citeN{MarkoffFalckeFender2001} the particles would be
accelerated up to a maximum energy $E_{\rm e,
max}=\gamma_{e,max}m_{\rm e}c^2$, which is reached when the
synchrotron loss rate equals that of acceleration.
We use the simple parallel shock acceleration rate
\begin{equation}
t_{\rm
acc}^{-1}=\frac{3}{4}\left(\frac{u_{\rm
sh}}{c}\right)^2\frac{eB}{m_{\rm e}c \xi
\gamma_{\rm e}},
\end{equation}
where $u_{\rm sh}$ is the shock speed in the plasma frame. The
parameter $\xi< c \beta_e/u_{\rm sh}$ \cite{Jokipii1987} is the ratio
between the parallel diffusive scattering mean free path and the
gyroradius of the particle, and ranges from a lower limit at $\xi=1$
up to typically a few $10^2$ (e.g., \citeNP{Jokipii1987}). For a
magnetic field of $\sim 20$ G as found in our model of the quiescent
state, the acceleration time scale is $\sim0.1$ sec for even
$\gamma_{\rm e}=10^5$ electrons, and hence is shorter than the
dynamical time scale at the black hole.

Setting the standard synchrotron loss rate $t_{\rm
syn}^{-1}=\frac{4}{3}\sigma_{\rm T} \gamma_{\rm e} \beta_{\rm e}^2
\frac{U_{\rm B}}{m_{\rm e} c}=t_{\rm acc}^{-1}$, we can solve for the
maximum electron energy achieved by acceleration $\gamma_{\rm
e,max}$.  If we define as a reference value $\xi=\xi_2 100$, the
maximum synchrotron frequency is then
\begin{equation}
\nu_{\rm max}=0.29 \nu_{\rm c} \simeq 1.2\cdot 10^{20} \xi_2^{-1}
\left(\frac{u_{\rm sh}}{c}\right)^2 \;\;\; {\rm Hz}
\end{equation}
where $\nu_{\rm c}\simeq\frac{3}{4\pi} \gamma_{\rm e,max}^2
(eB)/(m_{\rm e} c)$ is the critical synchrotron frequency.  This value
is not dependent on the magnetic field, the jet power, or the shock
location as long as we are in the synchrotron cooling dominated
regime.

Because the shock accelerated particles responsible for the X-ray
synchrotron will have very high energies ( $\gamma_{\rm e}\sim10^5$)
for the low magnetic fields further out in the jet, the synchrotron
cooling time scale will be very short, on the order of $\sim 10^2$ s.  This
means that re-acceleration along the jet is required to maintain the
population, and will result in rapid cooling if the acceleration is
switched off.  We thus approximate the shock acceleration as
continuous starting at a distance $z_{\rm sh}$.  For X-ray binaries we
found that the shock acceleration must begin relatively close to the
nozzle at $z_{\rm sh} \sim 10\sim 10^2 r_{\rm s}$ 
(\citeNP{MarkoffFalckeFender2001}; Markoff et al., in prep.).  This
location is determined from the data by extrapolating the synchrotron
X-ray curve to where it meets the optically-thick, flattish spectrum
in the radio-IR.  This intersection is unique for a fixed spectral
index, and gives $z_{\rm sh}$ because the self-absorption frequency
scales inversely with $z$ in the jet model.

\begin{figure}
\centerline{\psfig{figure=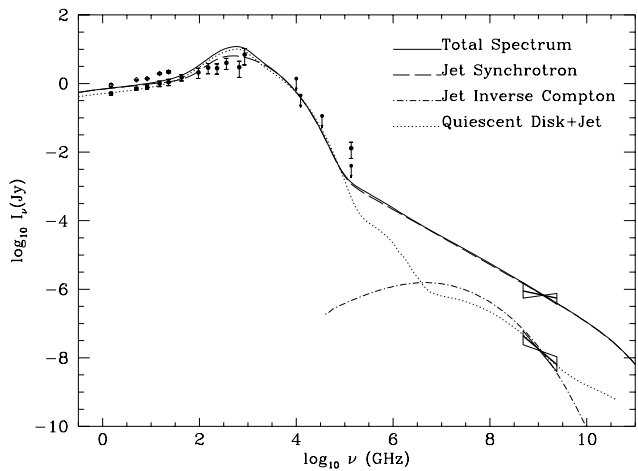,width=0.49\textwidth,angle=-90}}
\caption[]{\label{sgra_syn} Fit to the flare data of
\citeN{Baganoffetal2001b} for the shock-flare model, other data the
same as Fig.~1.}
\end{figure}

If we then fix for simplicity the fraction of accelerated particles at
50\% and keep the other parameters as in FM00 and
\citeN{YuanMarkoffFalcke2001} we can calculate the resultant
shock-flare model spectrum as shown in Fig.~\ref{sgra_syn}. As the
spectral index becomes harder for a fixed X-ray flux, the optically
thick turnover must occur at lower frequencies, i.e. further out in
the jet. For a standard spectral index of $\alpha\sim0.8$ as typically
seen in AGN, the shock acceleration region must be at $\sim 16 r_{\rm
s}$, which is consistent with the observed time scales. However, the
assumed standard AGN spectral index is only marginally compatible with
the spectrum observed for the X-ray flare, which poses a problem for
such a model.  Taking on the other hand the reported best-fit X-ray
spectral index at face value would imply $\alpha=0.3$ and require
$z_{\rm sh}\sim10^4 r_{\rm s}$. This is very far in comparison to
other jet systems and furthermore ruled out by the observed short time
scales.

\section{Discussion}

We are able to explain the 10 ks flare in Sgr A* detected by {\em
Chandra} by heating the radiating electrons within the jet model of
FM00, either so they remain quasi-thermalized ($T_{\rm e}$-flare) or
via the non-thermal process of shock acceleration (shock-flare). A
flare due to an increase in accretion rate ($\dot{M}$-flare) appears
very unlikely because of the generally low level of radio variability.

Of the two remaining scenarios, the shock-flare is more intriguing
because it offers a solution where the X-ray flare occurs without a
great effect on the lower-frequency emission, consistent with the
observed lower radio variability of Sgr A* over the last decades.  At
the marginal end of the fit, it can also explain the shortest
variations via the location of the shock or fast radiative cooling,
and predicts a spectral index consistent with that seen in other AGN
systems.  Although the radio flux does not change much for this case,
the presence of the optically thin tail would predict a significantly
larger radio profile (more extended, optically thin jet emission; see
FM00 for a discussion of this point) on the sky and a shift of the
centroid of the radio emission.  However, in the radio astrometric
work of \citeN{ReidReadheadVermeulen1999} no such shift has been
detected so far.

Alternatively, the $T_{\rm e}$-flare with its sudden heating of hot
($T_{\rm e}\simeq6\cdot10^{11}$ K) electrons by, e.g., magnetic
reconnection, can explain the X-ray flare via increased SSC emission,
similar to models for the quiescent state spectrum.  The fast
variability can be explained by the small source size and outflow with
$v\sim c$, leading to fast adiabatic cooling, while radiative cooling
is not as important ($t_{\rm syn}\sim5\cdot10^4$ for $T_{\rm
e}=6\cdot10^{11}$ K electrons in the submm-bump).  In contrast to the
shock-flare model, the $T_{\rm e}$-flare model fits the reported X-ray
spectrum much better.  However, the radio variability is larger than
in the synchrotron case, but still falls along the ``envelope'' of
highest radio fluxes observed so far (Fig. \ref{sgra_ssc}).  In
addition the model predicts simultaneous MIR flaring, in a regime where
no monitoring data is currently available.  So although the $T_{\rm
e}$-flare case is favored over the shock-flare case in terms of the
fit to the X-ray flare data, only simultaneous submm/MIR/X-ray and VLBI
observations in the near future will unequivocally determine its
viability.

For the $T_{\rm e}$-flare, assuming the protons have at least the same
temperature one can compare the energy density of the plasma and the
gravitational binding energy, $G M_\bullet m_{\rm p} n/R$, yielding
\begin{equation}
{\Gamma n k_{\rm b}T\over G M_\bullet m_{\rm p} n
R^{-1}}\simeq3.4\left({\tau\over 600 {\rm sec}}\right)\left({T\over
6\cdot10^{11} {\rm K}}\right)
\end{equation}
for a $M_\bullet=2.6\cdot10^6M_\odot$ black hole and a relativistic
plasma with $\Gamma=4/3$ (e.g., \citeNP{Konigl1980}) at a distance
$R=c\tau$ from the black hole set by the variability time
scale. Under such simple assumptions the plasma would not be
gravitationally bound, which is certainly consistent with the jet scenario.

\bibliography{aamnemonic,Di061}
\bibliographystyle{aa}

\end{document}